\documentclass[aps,pre,twocolumn,amsmath,amssymb,amsfonts,floatfix,superscriptaddress,showkeys,showpacs]{revtex4-1}
\usepackage{dcolumn}  
\usepackage{multirow}
\usepackage[T1]{fontenc}
\usepackage[]{inputenc}
\usepackage{dsfont}
\usepackage{verbatim}
\usepackage{bm} 
\usepackage{hyperref} 
\usepackage{cleveref} 
\usepackage{floatrow}
\usepackage{natbib}
\usepackage{graphicx}
\usepackage{epstopdf}
\usepackage{ifpdf} 
\usepackage{epsfig}
\usepackage{color}
\usepackage[usenames,dvipsnames]{xcolor}
\usepackage[caption=false]{subfig}
\usepackage[export]{adjustbox}
\usepackage[colorinlistoftodos]{todonotes}

\newcommand{\rmd}{{\mathrm{d}}}
\newcommand{\rme}{{\mathrm{e}}}
\newcommand{\icomplex}{\dot\iota}
\newcommand{\kBT}{k_{\mathrm{B}}T}
\newcommand{\kB}{k_{\mathrm{B}}}

\begin{document}

\title{Pseudo-Casimir interactions and surface anchoring duality in bookshelf geometry of smectic-A liquid crystals}

\author{Fahimeh \surname{Karimi Pour Haddadan}}\thanks{f.karimi@khu.ac.ir}
\affiliation{Faculty of Physics, Kharazmi University, Tehran 15815-3587, Iran}
	
\author{Ali \surname{Naji}}
\affiliation{School of Physics, Institute for Research in Fundamental Sciences (IPM), Tehran 19395-5531, Iran}

\author{Rudolf \surname{Podgornik}}
\affiliation{School of Physical Sciences and Kavli Institute for Theoretical Sciences, University of Chinese Academy of Sciences, Beijing 100049, China}
\affiliation{CAS Key Laboratory of Soft Matter Physics, Institute of Physics, Chinese Academy of Sciences (CAS), Beijing 100190, China}
\affiliation{Department of Physics, Faculty of Mathematics and Physics, University of Ljubljana, and
Department of Theoretical Physics, J. Stefan Institute, 1000 Ljubljana, Slovenia}

\begin{abstract}
We analyze the transverse intersubstrate pseudo-Casimir force, arising as a result of thermal fluctuations of the liquid crystalline layers of a smectic-A film confined between two planar substrates in a bookshelf geometry, in which the equidistant smectic layers are placed perpendicular to the bounding surfaces. We discuss the variation of the interaction force as a function of the intersubstrate separation in the presence of surface anchoring to the substrates, showing that the force induced by 
confined fluctuations is attractive and depends on the penetration length as well as the layer spacing. The strongest effect occurs for tightly confined fluctuations, in which the surface anchoring energy is set to infinity, where the force per area scales linearly with the thermal energy 
and inversely with the fourth power of the intersubstrate separation. By reducing  the strength of the surface anchoring energy, the force first becomes weaker in magnitude but then increases in magnitude as the surface anchoring strength is further reduced down to zero, in which case the force tends to that obtained in the limit of strong anchoring.
\end{abstract}


\maketitle

\section{Introduction}

Liquid crystals ~\cite{de-gennes} are soft materials characterized by long range, correlated thermal fluctuations of their order parameter. The external inclusions in these media, perturbing their ordered state, are thus expected to experience long-range fluctuation-induced forces akin to the van der Waals-Casimir force, which is induced between dielectric bodies by thermal as well as quantum fluctuations of the electromagnetic field inside and in the environment of the interacting bodies 
~\cite{Casimir1,Casimir2,Mostepanenko}. While this standard fluctuation-induced force also exhibits a thermal component, which is the zero-Matsubara-frequency term, the thermal fluctuation-induced interactions in the context of soft-matter systems came to be referred more generally as the {\em pseudo-Casimir force}. Such fluctuation-induced effects were studied extensively in various contexts of the soft materials such as liquid crystals or gels~\cite{LC2,LC3,rudi1,rudi2,rudi3,rudi4,ziherl,karimi1, karimi2,markun0,markun,Lyra1,Lyra2,golestanian,faar,fanr,haddadan1,haddadan2,bing1}, specifically also in the case of  the smectic-A phase~\cite{markun,Lyra1}. In this latter instance, the molecules are arranged in a series of parallel, equidistant, liquidlike smectic layers. While the molecules exhibit no particular positional order within each layer, they maintain a unidirectional orientational order with their long axes nearly aligned in the direction perpendicular to the plane of the smectic layers~\cite{de-gennes}.  To the lowest quadratic order, the mesoscopic free energy of the system is composed of the deformation energy of the orientational distortions within the layers as well as the compression energy of changing the separation between the layers~\cite{de-gennes}. Since both of these contributions depend on the square of the respective second and first order derivatives of the local layer displacement, the system should exhibit critical behavior characterized by long range correlations  ~\cite{golestanian}

In the present work, we focus on the pseudo-Casimir force produced between two plane-parallel walls (substrates) confining a smectic-A film in a so-called {\em bookshelf geometry}~\cite{LC2,radzihovsky1,radzihovsky2}, in which the bounding substrates are perpendicular to the fluctuating smectic-A layers, complementary to the rather different case of {\em paper-stack geometry}~\cite{Lyra1,Lyra2}, relevant in the context of the {\em vapor pressure paradox} for confined lipid multilayers ~\cite{rudivapor,leonardo}. Thermal fluctuations generate local distortions within this one dimensional layered structure, which in turn give rise to a fluctuation-induced force between the substrates of the smectic film, mitigated by the finite energy penalty for the deformation at the bounding surfaces. The bookshelf geometry has been briefly considered in Ref.~\cite{LC2}, where by ``excluded particle'' argument the force per area is proposed to be proportional to $-\kBT \lambda/d^4$, where $\kB$ is the Boltzmann constant, $T$ is the temperature, $\lambda$ is a characteristic length and $d$ is the intersubstrate separation. In distinction, here we derive the fluctuational force by direct and systematic  calculations,  considering explicitly the undulations of the smectic layers bounded by hard surfaces to which the layer displacement is coupled by a finite surface anchoring energy. We assume that the fluctuational modes of the smectic layers are equally attenuated by identical surface-anchoring fields at both planar boundaries. We shall however ignore the fluctuations in the degree of smectic order and the possible couplings  between the layer undulations and the director field fluctuations, describing the orientational order of the molecules within the smectic phase. The role of such a coupling has been studied in the case of a homeotropic and a free-standing smectic film  in Ref.~\cite{markun}.

We introduce our model in Section~\ref{sec:model} and discuss the functional-integral methods used to evaluate the partition function and, hence, the free energy of the system in Section \ref{sec:formalism}, where the analytical and numerical results are discussed in detail (Sections \ref{sec:force} and \ref{sec:force_num}) with the concluding remarks to be followed in Section \ref{sec:con}.

\begin{figure}
\begin{center}
\includegraphics[width=10cm]{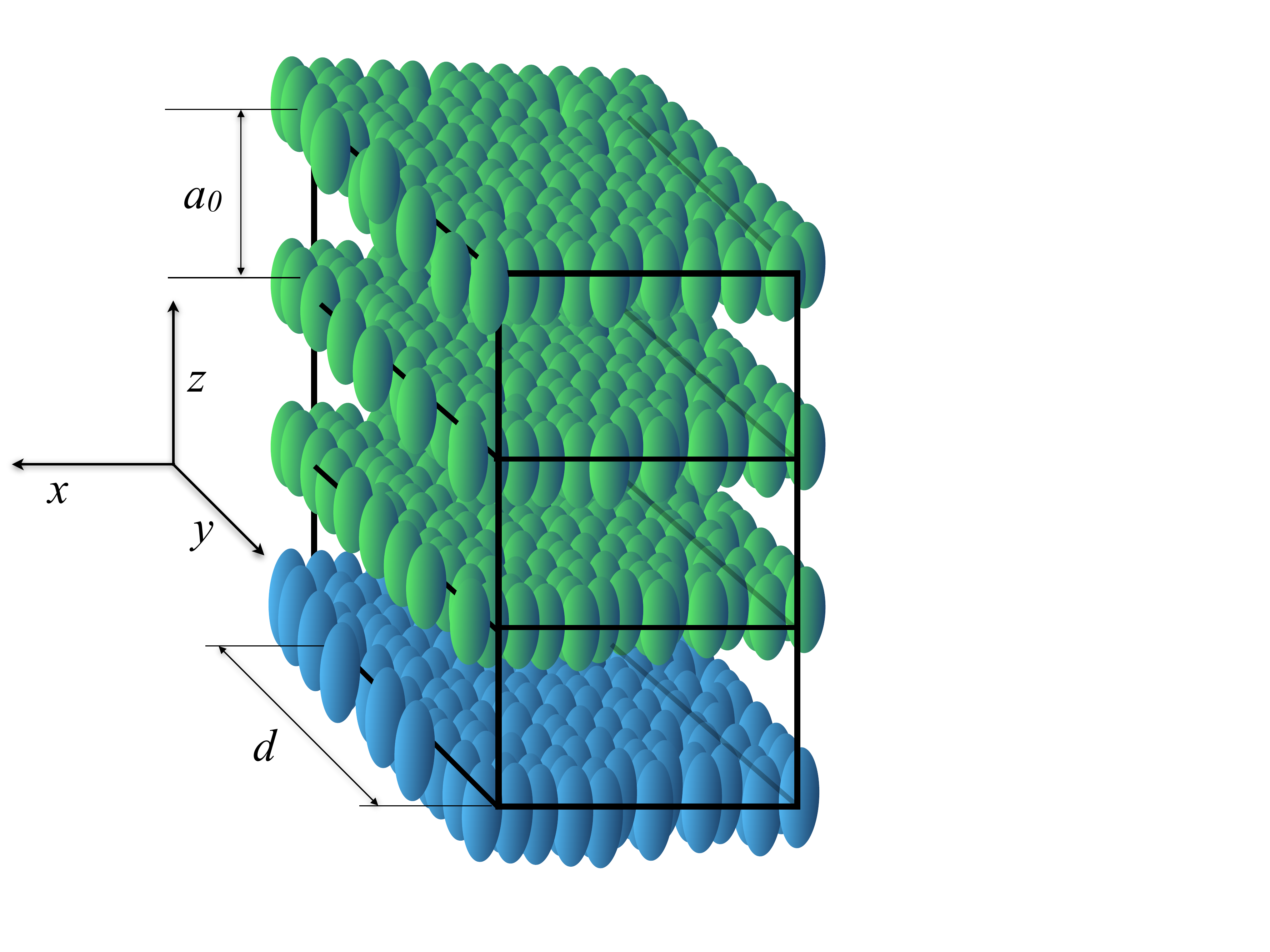}
\caption{
A schematic presentation of the ground-state `bookshelf' configuration of a smectic-A film confined between planar side substrates of separation $d$ along the $y$-axis. The smectic layer spacing, $a_0$, along the $z$-axis is around the molecular length ($a_0$ is not drawn to scale). The system is unbounded in both $x$ and $z$ directions.}
\label{schematic}
\end{center}
\end{figure}

\section{Model}
\label{sec:model}

We consider a smectic-A phase of a liquid-crystalline material~\cite{de-gennes} with parallel layers of separation  $a_0$, stacked up in the $z$ direction. The long axes of the molecules, comprising each of the liquidlike smectic layers, are thus aligned on average in the $z$ (normal-to-layer) direction. The smectic-A phase is assumed to be confined between two plane-parallel walls (substrates) placed perpendicular to the smectic layers at positions $y=0$ and $d$; see Fig. \ref{schematic}. The displacement field due to thermal (in-plane and out-of-plane) undulations of the smectic layers is denoted by  $u({\mathbf r}_{\!_\perp}, y)$, while the director field of the molecules is denoted by ${\mathbf n}({\mathbf r}_{\!_\perp}, y)$, where ${\mathbf r}=({\mathbf r}_{\!_\perp}, y)$ is the Cartesian spatial position with the transverse coordinates denoted by ${\mathbf r}_{\!_\perp}=(x,z)$. 

The Hamiltonian (deformation energy) associated with such displacements is given by 
\begin{equation}
{\cal H}_b=\frac{K}{2}\int_{V} \rmd{\mathbf r} \left[\left(\partial_{x}^2 u+\partial_{y}^2 u\right)^2+{B\over K}(\partial_{z}u)^2\right],
\label{bulk}
\end{equation}
where $K$ and $B$ are elastic moduli corresponding to bending and compressing (or, dilating) of the layers, respectively, and $V$ is the volume of the smectic slab~\cite{de-gennes}.  

We assume in general that a preferred order exists for the director field ${\mathbf n}$ at the confining substrates of the smectic film, that is, ${\mathbf n}({\mathbf r}_{\!_\perp},0)$ and ${\mathbf n}({\mathbf r}_{\!_\perp},d)$ orient preferentially along the $z$-axis. The latter is modeled using the Rapini-Papoular model~\cite{RP} by assuming the surface-interaction Hamiltonian 
\begin{equation}
{\cal H}_s=-\frac{W}{2}\int_{\partial V} \rmd{\mathbf r}_{\!_\perp}\left({\mathbf n}\cdot \hat{\mathbf z}\right)^2,
\end{equation}
where $W$ is the anchoring energy per unit area and the integration is taken over  the surface area of the confining  boundaries (substrates)~\cite{radzihovsky1}. Considering small thermal fluctuations of the director field around its ground-state configuration, that is, by writing ${\mathbf n} = \hat{\mathbf z}+\delta {\mathbf n}=(\delta n_{x},\delta n_{y},\sqrt{1-\delta n_{x}^2-\delta n_{y}^2})$, we find
\begin{equation}
{\cal H}_s=\frac{W}{2}\int_{\partial V} \rmd{\mathbf r}_{\!_\perp}\left[(\partial_{x}u)^2+(\partial_{y}u)^2 \right]
\label{third}
\end{equation}
up to the quadratic order in the fluctuating field $\delta {\mathbf n}=(\delta n_{x}, \delta n_y)$, where we have also used the relation $(\partial_{x},\partial_{y})^T u+\delta {\mathbf n}=0$,  asserting that the local director field remains perpendicular to the smectic layers~\cite{de-gennes,radzihovsky1,lubensky}.

Because of the translational invariance of the Hamiltonian across the $(x,z)$-plane, we can use the Fourier representation $u({\mathbf r}_{\!_\perp},y)=\sum_{{\mathbf p}_{\!_\perp}}u_{{\mathbf p}_{\!_\perp}}\!(y)\,\rme^{\icomplex {\mathbf p}_{\!_\perp}\!\cdot{\mathbf r}_{\!_\perp} }$, with ${\mathbf p}_{\!_\perp}=(p_x, p_z)$ being the two-dimensional wavevector, to express the total Hamiltonian ${\cal H}={\cal H}_b+{\cal H}_s$ in terms of the terms contributed by individual transverse modes ${\mathbf p}_{\!_\perp}$. Hence, the total Hamiltonian as well as its bulk and surface parts can be additively decomposed as ${\cal H} \equiv \sum_{{\mathbf p}_{\!_\perp}}h\big[u_{{\mathbf p}_{\!_\perp}}\!(y)\big]$, into a bulk functional ${\cal H}_b \equiv \sum_{{\mathbf p}_{\!_\perp}}h_b\big[u_{{\mathbf p}_{\!_\perp}}\!(y)\big]$ and a surface function ${\cal H}_s \equiv \sum_{{\mathbf p}_{\!_\perp}}h_s\big(u_{{\mathbf p}_{\!_\perp}}\!(0), u_{{\mathbf p}_{\!_\perp}}\!(d)\big)$, where obviously 
\begin{equation}
h\big[u_{{\mathbf p}_{\!_\perp}}\!(y)\big] =  h_b\big[u_{{\mathbf p}_{\!_\perp}}\!(y)\big]+h_s\big(u_{{\mathbf p}_{\!_\perp}}\!(0), u_{{\mathbf p}_{\!_\perp}}\!(d)\big), 
\end{equation}
and the bulk and surface terms associated with each transverse mode are obtained, respectively, as
\begin{eqnarray}
\label{ham_b}
h_b\big[u_{{\mathbf p}_{\!_\perp}}\!(y)\big]&&\,=\frac{KA}{2}\!\int_{0}^{d}\! \rmd y\, \Big\{ |\ddot u_{{\mathbf p}_{\!_\perp}}\!(y)|^2 + 2 p_x^2 |\dot u_{{\mathbf p}_{\!_\perp}}\!(y)|^2 \nonumber\\
&&\,+\, (p_x^4 + \lambda^{-2} p_z^2)|u_{{\mathbf p}_{\!_\perp}}\!(y)|^2\Big\},
\end{eqnarray}
\begin{eqnarray}
\label{ham_s}
h_s\big(&&u_{{\mathbf p}_{\!_\perp}}\!(0),u_{{\mathbf p}_{\!_\perp}}\!(d)\big) \,= \frac{WA}{2} \Big\{ |\dot u_{{\mathbf p}_{\!_\perp}}\!(0)|^2 + |\dot u_{{\mathbf p}_{\!_\perp}}\!(d)|^2\nonumber\\
&&\,+\, p_x^2\big(|u_{{\mathbf p}_{\!_\perp}}\!(0)|^2+|u_{{\mathbf p}_{\!_\perp}}\!(d)|^2\big)\Big\} 
\\
&&
-\,\frac{KA}{2} p_x^2 \Big\{\big(u_{{\mathbf p}_{\!_\perp}}\!(d) \dot u_{{\mathbf p}_{\!_\perp}}^{\ast}(d)-u_{{\mathbf p}_{\!_\perp}}\!(0) \dot u_{{\mathbf p}_{\!_\perp}}^{\ast}(0)\big)+c.c.\Big\},
\nonumber
\end{eqnarray}
where the penetration length is defined as $\lambda=\sqrt{K/B}$~\cite{de-gennes}, $A$ is the surface area of each confining plate,
$u_{{\mathbf p}_{\!_\perp}}^{*}(y)=u_{-{\mathbf p}_{\!_\perp}}(y)$, and the dots over the symbols denote the $y$-derivative, i.e., ${\dot u}_{{\mathbf p}_{\!_\perp}}\equiv \partial_{y}u_{{\mathbf p}_{\!_\perp}}$. 
\section{Formalism and results}
\label{sec:formalism}

\subsection{Partition function}

The partition function of the system described in the previous section follows by integrating over all field fluctuations as
\begin{equation}
{\cal Z}=\prod_{{\mathbf p}_{\!_\perp}}\int {\cal D}u_{{\mathbf p}_{\!_\perp}}\!(y)\,\rme^{-\beta h[u_{{\mathbf p}_{\!_\perp}}\!(y)]}\equiv \prod_{{\mathbf p}_{\!_\perp}}\rme^{-\beta {\cal F}({\mathbf p}_{\!_\perp})}, 
\label{gfnchjkw}
\end{equation}
where $\beta=1/(\kBT)$ and ${\cal F}({\mathbf p}_{\!_\perp})$ is the free energy contributed by each transverse mode ${\mathbf p}_{\!_\perp}$. The thermodynamic free energy, ${\cal F} = -\kBT\ln {\cal Z} = \sum_{{\mathbf p}_{\!_\perp}} {\cal F}({\mathbf p}_{\!_\perp})$, follows in the continuum limit  as 
\begin{equation} 
{\cal F}(d)=  \frac{A}{2\pi^2}\int_{-\infty}^\infty \rmd p_x \!\int_{0}^{\pi/a_0}\rmd p_z ~{\cal F}(
p_x,p_z),
\end{equation}
where we have explicitly noted the dependence of the free energy on the intersubstrate distance, $d$, and taken the $p_z$ integral in the first Brillouin zone $-\pi/a_0\leq p_z<\pi/a_0$. 

The part of the partition function corresponding to the bulk Hamiltonian, itself a second derivative functional of the fluctuating fields,  can be evaluated explicitly for any configuration of the surface fields, $u_{{\mathbf p}_{\!_\perp}}\!(0), \dot u_{{\mathbf p}_{\!_\perp}}\!(0),  u_{{\mathbf p}_{\!_\perp}}\!(d)$, and $\dot u_{{\mathbf p}_{\!_\perp}}\!(d)$~\cite{kleinert}, which finally leads to a compact epression for the full partition function Eq. \eqref{gfnchjkw} in the form
\begin{widetext}
\begin{eqnarray}
{\cal Z}= \prod_{{\mathbf p}_{\!_\perp}} \int \rmd(u_{{\mathbf p}_{\!_\perp}}\!(0))\rmd(\dot u_{{\mathbf p}_{\!_\perp}}\!(0)) 
\rmd(u_{{\mathbf p}_{\!_\perp}}\!(d))\rmd(\dot u_{{\mathbf p}_{\!_\perp}}\!(d))  \, \rme^{- \beta h_s(u_{{\mathbf p}_{\!_\perp}}\!(0))} {\cal G}\Big( \dot u_{{\mathbf p}_{\!_\perp}}\!(0), u_{{\mathbf p}_{\!_\perp}}\!(0); \dot u_{{\mathbf p}_{\!_\perp}}\!(d), u_{{\mathbf p}_{\!_\perp}}\!(d)\Big) \rme^{- \beta h_s(u_{{\mathbf p}_{\!_\perp}}\!(d))},
\label{partition_fuction}
\end{eqnarray}
\end{widetext}
where the propagator ${\cal G}\Big( \dot u_{{\mathbf p}_{\!_\perp}}\!(0), u_{{\mathbf p}_{\!_\perp}}\!(0); \dot u_{{\mathbf p}_{\!_\perp}}\!(d), u_{{\mathbf p}_{\!_\perp}}\!(d)\Big)$, giving the statistical weight of bounding surfaces field configurations characterized by $\!\dot u_{{\mathbf p}_{\!_\perp}}\!(0), u_{{\mathbf p}_{\!_\perp}}\!(0)$ and $ \dot u_{{\mathbf p}_{\!_\perp}}\!(d), u_{{\mathbf p}_{\!_\perp}}\!(d)$, is defined as
\begin{equation}
{\cal G}\Big(\!\dot u_{{\mathbf p}_{\!_\perp}}\!(0), u_{{\mathbf p}_{\!_\perp}}\!(0); \dot u_{{\mathbf p}_{\!_\perp}}\!(d), u_{{\mathbf p}_{\!_\perp}}\!(d)\!\Big) \!=\!\! \!\int \!\! {\cal D}u_{{\mathbf p}_{\!_\perp}}\!(y) \,\rme^{- \beta h_b[u_{{\mathbf p}_{\!_\perp}}\!(y)]}.
\end{equation}
Note that the surface term \eqref{ham_s} is additive in $u_{{\mathbf p}_{\!_\perp}}\!(0)$ and $u_{{\mathbf p}_{\!_\perp}}\!(d)$ and, as such, it is routinely grouped into two separate terms, denoted by $h_s\big(u_{{\mathbf p}_{\!_\perp}}\!(0)\big)$ and $h_s\big(u_{{\mathbf p}_{\!_\perp}}\!(d)\big)$.
The propagator ${\cal G}$ involves functional integration over a Boltzmann-wighted Hamiltonian involving the second-order derivative of the field $u_{{\mathbf p}_{\!_\perp}}\!(y)$; see Eq.~(\ref{bulk}), which can be calculated using the methods introduced  in Ref.~\cite{kleinert}, or the final expression provided in Ref. \cite{handbook}. As a matter of fact, this methodology has already been used in the context of Casimir interactions, e.g., in Ref.~\cite{dobnikar}, dealing with the pseudo- Casimir force in a confined nematic polymer, and in Ref.~\cite{uchida}, which analyzes the Casimir effect in fluids above the isotropic-lamellar transition within the Brazovskii mesoscopic theory, with formal developments discussed further in Ref. \cite{dean}. Hence, using the same methodology and basically following Ref.~\cite{kleinert}, we derive a closed form expression for the integrand entering the partition function \eqref{partition_fuction} as 
\begin{widetext}
\begin{eqnarray}
\rme^{- \beta h_s(u_{{\mathbf p}_{\!_\perp}}\!(0))}{\cal G}\Big( \dot u_{{\mathbf p}_{\!_\perp}}\!(0), u_{{\mathbf p}_{\!_\perp}}\!(0); \dot u_{{\mathbf p}_{\!_\perp}}\!(d), u_{{\mathbf p}_{\!_\perp}}\!(d)\Big) \rme^{- \beta h_s(u_{{\mathbf p}_{\!_\perp}}\!(d))} = {\cal A}(\omega_1, \omega_2, d) \, \rme^{-\beta h_s^{\mathrm{eff}}\big(\dot u_{{\mathbf p}_{\!_\perp}}\!(0), u_{{\mathbf p}_{\!_\perp}}\!(0); \dot u_{{\mathbf p}_{\!_\perp}}\!(d), u_{{\mathbf p}_{\!_\perp}}\!(d)\big)} 
\label{eq:A_h_s_eff}
\end{eqnarray}
\end{widetext}
with an effective field surface-Hamiltonian that can be obtained as
\begin{eqnarray}
&&h_s^{\mathrm{eff}}\big(u_{{\mathbf p}_{\!_\perp}}\!(0), u_{{\mathbf p}_{\!_\perp}}\!(d); \dot u_{{\mathbf p}_{\!_\perp}}\!(0), \dot u_{{\mathbf p}_{\!_\perp}}\!(d) \big)\nonumber\\ 
&&\qquad\qquad\qquad =\,m_{11} \big(|u_{{\mathbf p}_{\!_\perp}}\!(0)|^2 + u_{{\mathbf p}_{\!_\perp}}\!|(d)|^2\big) \nonumber\\ 
&&\qquad\qquad\qquad+\,m_{12}\big(u_{{\mathbf p}_{\!_\perp}}\!(0)u_{{\mathbf p}_{\!_\perp}}\!^{\ast}(d)+c.c.\big)\nonumber\\ 
&&\qquad\qquad\qquad+\,m_{13} \big(\dot u_{{\mathbf p}_{\!_\perp}}\!(d) u^{\ast}(d) - \dot u_{{\mathbf p}_{\!_\perp}}\!(0) u_{{\mathbf p}_{\!_\perp}}\!^{\ast}(0)+c.c.\big)\nonumber\\ 
&&\qquad\qquad\qquad+ \,m_{14} \big(\dot u(d) u^{\ast}(0) - \dot u(0) u^{\ast}(d)+c.c.\big)\nonumber\\ 
&&\qquad\qquad\qquad+\,m_{33} \big(|\dot u(0)|^2 + |\dot u(d)|^2\big) \nonumber\\ 
&&\qquad\qquad\qquad+ \,m_{34} \big(\dot u(0)\dot u^{\ast}(d)+c.c\big).
\label{eq:h_s_eff}
\end{eqnarray}
We make use of the eigenfrequencies $\omega_1 $ and $\omega_2$ [see Eq.~\eqref{ham_b}] defined through 
\begin{eqnarray}
&&\omega_1 ^2+\omega_2 ^2=2 p_x ^2,\nonumber\\
&&\omega_1 ^2\omega_2 ^2=p_x ^4+\lambda^{-2}p_z^2,
\label{omega}
\end{eqnarray}
where again $\lambda=\sqrt{K/B}$ is the penetration length, $\ell=K/W$ is the extrapolation (or pinning) length, and the shorthand notations $c_i\equiv\cosh (\omega_i d) $ and $s_i\equiv\sinh (\omega_i d)$ for $i=1,2$~\cite{kleinert} to express  the elements  $m_{ik}$ of the matrix $\hat m$ appearing in Eq. \eqref{eq:h_s_eff} as
\begin{eqnarray}
m_{11}&=&\frac{\omega_1 \omega_2}{2M} (\omega_{1}^2-\omega_{2}^2)(\omega_{1} s_{1}c_{2}-\omega_{2} c_{1} s_{2})
+{p_x^2\over 2\ell}
\\
m_{12}&=&-\frac{\omega_1 \omega_2}{2M} (\omega_{1}^2-\omega_{2}^2)(\omega_{1} s_{1}-\omega_{2} s_{2})
\\
m_{13}&=&\frac{\omega_1 \omega_2}{2M} [(\omega_{1}^2+\omega_{2}^2)(c_1 c_2 -1)+2\omega_1 \omega_2 s_1 s_2]
+{p_x^2\over 2}
\nonumber\\
\\
m_{14}&=&-\frac{\omega_1 \omega_2}{2M}(\omega_{1}^2-\omega_{2}^2)(c_1 -c_2)
\\
m_{33}&=&{1\over 2M}(\omega_{1}^2-\omega_{2}^2)(\omega_{1} s_{2}c_{1}-\omega_{2} c_{2} s_{1})
+{1\over 2\ell}
\\
m_{34}&=&-{1\over 2M}(\omega_{1}^2-\omega_{2}^2)(\omega_1 s_2-\omega_2 s_1), 
\label{defmik}
\end{eqnarray}
and the prefactor ${\cal A}$ in Eq. \eqref{eq:A_h_s_eff} as 
\begin{equation}
 {\cal A}(\omega_1, \omega_2, d)={1\over 2\pi}{\sqrt{\omega_1 \omega_2}|\omega_1^2-\omega_2^2|\over \sqrt{M}}.
\label{F_beta}
\end{equation}
where 
\begin{equation} 
M=(\omega_1^2+\omega_2^2)s_1 s_2-2\omega_1 \omega_2 (c_1 c_2 -1). 
\label{detM}
\end{equation}
The final step of calculating ${\cal Z}$ involves Gaussian functional integrals over the fluctuating surface fields $\{\dot u_{{\mathbf p}_{\!_\perp}}\!(0), u_{{\mathbf p}_{\!_\perp}}\!(0); \dot u_{{\mathbf p}_{\!_\perp}}\!(d), u_{{\mathbf p}_{\!_\perp}}\!(d)\}$ of the factor $\rme^{-\beta h_s^{\mathrm{eff}}}$, see Eq. \eqref{eq:A_h_s_eff}.   

\subsection{Interaction free energy}
\label{subsec:free}

The free energy straightforwardly follows from the final results in the preceding section. By dropping irrelevant additive factors as necessary, the interaction free energy of the system can be 
obtained from 
\begin{equation} 
\label{free_energy_final}
{\cal F}(d)  =-k_B T \ln {\cal Z}, 
\end{equation}
where 
\begin{equation}
{\cal Z}\propto \prod_{{\mathbf p}_{\!_\perp}}
{1\over \sqrt{|G({\mathbf p}_{\!_\perp})|}},
\end{equation}
and $G({\mathbf p}_{\!_\perp})$ can be 
written in an explicit form as
\begin{eqnarray}
G({\mathbf p}_{\!_\perp})&&\,=A+B\cosh (2\alpha d)+C \sinh (2\alpha d)\nonumber\\
&&\,+\,D\cos (2\gamma d)+E\sin (2\gamma d).
\label{green}
\end{eqnarray}
Here $\alpha$ and $\gamma$ are defined through the real and imaginary parts of the solutions of  Eqs.~(\ref{omega}), as $\omega_1=\alpha-\icomplex\gamma$ and $\omega_2=\alpha+\icomplex\gamma$, where
\begin{eqnarray}
&&\alpha=\frac{1}{\sqrt{2}}\left(p_{x}^2+\sqrt{p_{x}^4+{\lambda^{-2}p_z^2}}\right)^{1/2},
\label{alpha}
\\
&&\gamma=\frac{1}{\sqrt{2}}\left(-p_{x}^2+\sqrt{p_{x}^4+{\lambda^{-2}p_z^2}}\right)^{1/2}, 
\label{gamma}
\end{eqnarray}
and we find 
\begin{eqnarray}
&&A=\sqrt{p_x^4 + \lambda^{-2}p_z^2}(\lambda^{-2}p_z^2 - \ell^{-2}p_x^2 )^2, \\
&&B=-\gamma^2 [\lambda^{-4}p_z^4 + 2\lambda^{-2} p_z^2 (4 p_x^2 + \sqrt{p_x^4 +\lambda^{-2} p_z^2}) \ell^{-2} \nonumber\\ 
&& \quad\quad\,+\,\ell^{-2} p_x^4  (8 p_x^2 + 8 \sqrt{p_x^4 +\lambda^{-2} p_z^2} + \ell^{-2})],\\
&&C=-2\alpha \lambda^{-2}p_z^2 \ell^{-1} (\lambda^{-2}p_z^2 +\ell^{-2} p_x^2 ), \\
&&D=-\alpha^2 [\lambda^{-4}p_z^4 -2\lambda^{-2} p_z^2 (-4 p_x^2 + \sqrt{p_x^4 + \lambda^{-2}p_z^2})\ell^{-2} \nonumber\\
&& \quad\quad\,+\, \ell^{-2} p_x^4  (8 p_x^2 - 8 \sqrt{p_x^4 +\lambda^{-2} p_z^2} + \ell^{-2})],\\
&&E=-2\gamma \lambda^{-2}p_z^2 \ell^{-1} (\lambda^{-2}p_z^2 + \ell^{-2}p_x^2 ).
\label{coefficients}
\end{eqnarray}

\subsection{Limiting cases}
\label{sec:force}

In the limits of weak and strong anchoring, i.e., $W\rightarrow 0$ and  $W\rightarrow\infty$, 
or, equivalently, expressed in terms of the extrapolation length, $\ell \rightarrow \infty$ and $\ell \rightarrow 0$, respectively, $G$ (Eqs.~(\ref{green})-(\ref{coefficients})), irrespective of irrelevent prefactors, reduces to the exact same expression  
\begin{eqnarray}
G\big|_{W\rightarrow 0,\infty}&&\,=\alpha^2+\gamma^2-\gamma^2\cosh (2\alpha d)-\alpha^2\cos (2\gamma d) \nonumber\\
&&\,=\gamma^2\sinh^2 (\alpha d)-\alpha^2\sin^2 (\gamma d), 
\label{G_limit}
\end{eqnarray}
which  is equal to $M/2$ [Eq.~(\ref{detM})].
In other words, the interaction free energy in the strong and the weak anchoring limits is found to coincide. This is a special case of a {\em duality} that exists in this system and tallies also with the properties of the pseudo-Casimir interactions in nematic liquid crystals~\cite{LC2,LC3}.

The interaction free energy per unit area (after subtracting a trivial additive contribution from the bulk material) is obtained as 
\begin{eqnarray}
\label{strong_anchoring}
{{\cal F}(d)\over A}&&\,= {\kBT \lambda\over  2\pi^2 d^3}\!\int_0^\infty\!\! \rmd p_x \int_0^\infty\!\!\rmd p_z 
\ln\Big\{(1-\rme^{-2\alpha})^2\\
&&\qquad\qquad\qquad\qquad\quad\times\Big[1-\Big({2\alpha \sin \gamma\over \gamma (\rme^{2 \alpha}-1)}\Big)^2 \rme^{2\alpha}\Big]\Big\},
\nonumber
\end{eqnarray}
where we have defined the rescaled variables  $p_x d\rightarrow p_x$, $p_z (d^2/\lambda)\rightarrow p_z$ and have taken the upper limit of $p_z$ to infinity, i.e., ${\pi d^2}/({\lambda a_0})\rightarrow \infty$. We have also non-dimensionalized $\alpha$ and $\gamma$ [Eqs. \eqref{alpha}, \eqref{gamma}] as 
\begin{eqnarray}
&&\alpha d\rightarrow \alpha =\big(q_x^2+\sqrt{q_x^4+q_z^2}\big)^{1/2}/\sqrt{2}, \\
&&\gamma d\rightarrow \gamma=\big(\!-q_x^2+\sqrt{q_x^4+q_z^2}\big)^{1/2}/\sqrt{2}.
\end{eqnarray}

\begin{figure}
\begin{center}
\includegraphics[width=7.7cm]{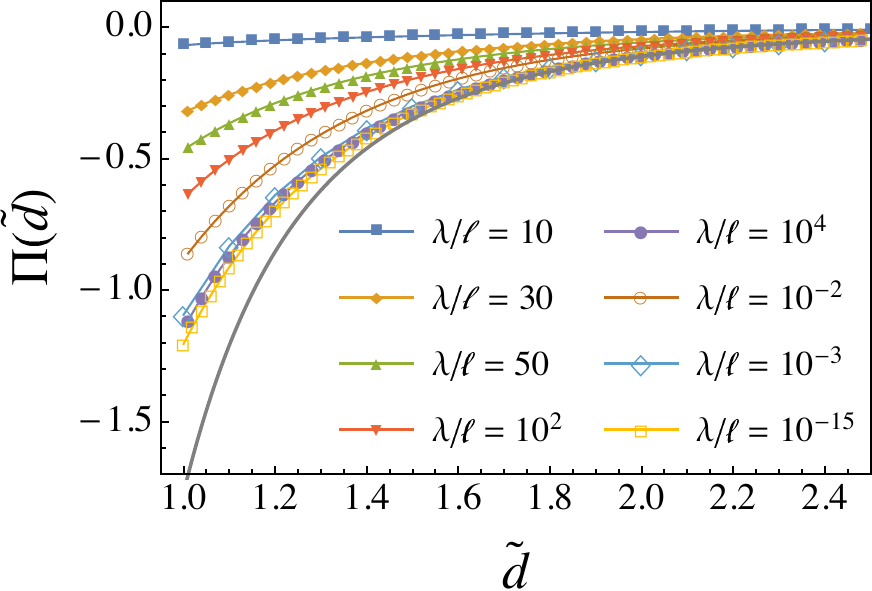}
\caption{Dimensionless interaction pressure ${\Pi}(\tilde d)$ as a function of the rescaled intersubstrate distance $\tilde d=d/\lambda$ for fixed $\lambda/a_0=2$ and $\lambda/\ell =10^{-15}, 10^{-3}, 10^{-2}$ (open symbols), and $10, 30, 50, 10^2$, and $ 10^4$ (filled symbols), as indicated on the graph. Curves connecting the symbols are guides to the eye. 
Analytical expression \eqref{eq:pressure_0} is shown as a gray solid curve. 
}
\label{fig:fig2}
\end{center}
\end{figure}

\subsection{Numerical results}
\label{sec:force_num}

\begin{figure}
	\begin{center}
		\includegraphics[width=7.7cm]{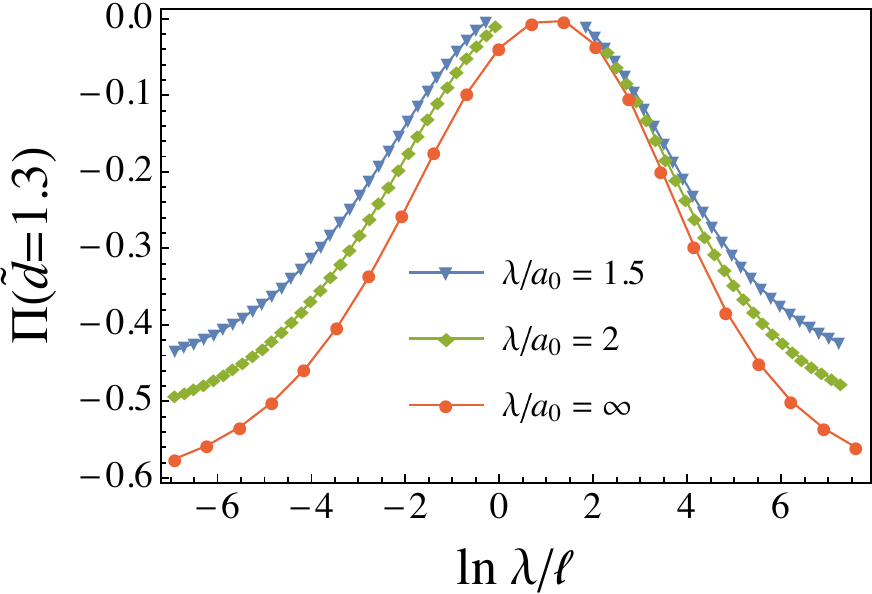}
		\caption{Dimensionless interaction pressure ${\Pi}(\tilde d)$ as a function of $\ln (\lambda/\ell)$  for fixed $\tilde d = 1.3$  and $\lambda/a_0=1.5, 2$, and $\infty$, as indicated on the graph. Curves are guides to the eye.
			For small enough $\lambda/a_0$ the pressure on the scale of the graph would be zero, though remaining negative.}
	    \label{fig:fig3}
	\end{center}
\end{figure}

\begin{figure}
	\begin{center}
		\includegraphics[width=7.7cm]{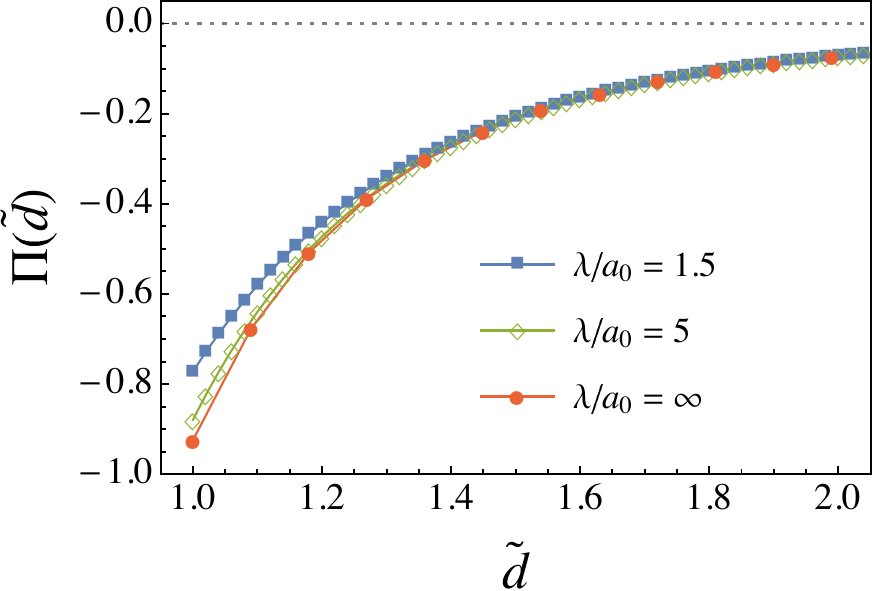}
		\caption{Dimensionless interaction pressure ${\Pi}(\tilde d)$ as a function of $\tilde d$  for fixed $\lambda/\ell=100$  and $\lambda/a_0=1.5, 5$, and $\infty$, as indicated on the graph. Curves connecting the symbols are guides to the eye.}		
		\label{fig:fig4}
	\end{center}
\end{figure}

The free energy \eqref{strong_anchoring} gives rise to a fluctuation-induced interaction force, 
\begin{equation} 
\label{eq:force}
f(d) = -\frac{\partial {\cal F}(d)}{\partial d}, 
\end{equation}
which turns out to be  attractive for the whole range of the intersubstrate separations. We numerically evaluate the corresponding dimensionless interaction pressure, 
\begin{equation} 
\label{eq:pressure}
\Pi(\tilde d)\equiv  \frac{\beta \lambda^3}{A}f(\lambda \tilde d), 
\end{equation}
 and show its behavior as a function of the rescaled intersubstrate separation $\tilde d=d/\lambda$ and other rescaled system parameters, i.e., $\lambda/\ell$ and $\lambda/a_0$, in Figs. \ref{fig:fig2}, \ref{fig:fig3} and \ref{fig:fig4}, respectively. To produce the numerical results, the model parameters are varied as follows. First, we note that $\lambda$ is typically of the order of $10^{-7}$~cm for $B=10^{8}$~erg/cm$^3$ and $K=10^{-6}$~dyn, and the layer thickness $a_0$ is expected to be around the molecular length of about $10^{-7}$~cm \cite{de-gennes}. In other words, $\lambda/a_0$ would be of order one. The anchoring extrapolation length $\ell$ can, on the other hand, be varied from zero to infinity. For completeness, however, we take a few different values for  the ratio $\lambda/a_0=1.5, 2$ and $\infty$, and a much wider range of values for $\lambda/\ell=10^{-15}-10^4$. 

Figure~\ref{fig:fig2} shows $\Pi(\tilde d)$ as a function of  $\tilde d$ for fixed $\lambda/a_0=2$ and for both large and small values of $\lambda/\ell$. The results are computed using Eqs. \eqref{free_energy_final}-\eqref{coefficients} and \eqref{eq:pressure}. Starting with a fixed $\lambda/\ell>1$ and increasing this ratio to very large values, the pressure profiles become increasingly less negative (strong anchoring). The same trend occurs starting with a fixed $\lambda/\ell<1$ and decreasing it to very small values (weak anchoring), corroborating the aforementioned observation that  the limiting results for $\lambda/\ell\rightarrow \infty$ and 0 coincide, where the largest magnitude of the attractive (negative) pseudo-Casimir force at a given rescaled separation $\tilde d$ is established. 

The interaction pressure in the strict limit of strong (or, weak) anchoring can be quite accurately approximated by a universal expression as
	\begin{equation} 
	\label{eq:pressure_0}
	{\Pi}_0(\tilde d)=  -C_0/\tilde d^4, 
	\end{equation}
where $C_0\simeq 1.76985$ is obtained from numerical integration of Eq. \eqref{strong_anchoring}. Equation \eqref{eq:pressure_0} is shown as a gray curve in Fig. \ref{fig:fig2} and, as seen, it closely approximates the data obtained for large (or, small) $\lambda/\ell$, when $\tilde d$ is only modestly large (i.e., $\tilde d\gtrsim 1.5$; the above analytical estimate is in fact expected to remain valid for $d^2\gg\lambda a_0$).

Figure~\ref{fig:fig2} also indicates that, at a given value of $\tilde d$ and for fixed $\lambda/a_0$, the interaction pressure varies {\em non-monotonically} with $\lambda/\ell$ (or, equivalently, with the anchoring energy). This behavior is shown in Fig.~\ref{fig:fig3}, where the dimensionless pressure is plotted as a function of $\ln \lambda/\ell$, at fixed $d/\lambda=1.3$. It turns out that the magnitude of the (attractive) interaction pressure increases as $\lambda/a_0$ is increased (or, equivalently, $a_0$ is decreased). This can be seen from both Figs.~\ref{fig:fig3} and ~\ref{fig:fig4}. In the latter figure, we fix $\lambda/\ell=100$ and plot the interaction pressure profiles for three different values of $\lambda/a_0$ , as indicated on the graph. 

\section{Concluding remarks}
\label{sec:con}

We considered a smectic-A liquid crystalline film confined within a planar gap that quenches the thermal fluctuations at the boundaries of the sample. A bookshelf geometry--in which the molecules vicinal to the two planar bounding substrates are forced to orient in co-planar directions--provides a particularly interesting physical system to study the fluctuating modes of the layering field and the layer displacement of the smectic-A phase~\cite{thesis}. The imposition of the hard boundaries modifies the free-space fluctuation spectrum, allowing only for the modes which are compatible with the presence of the boundaries as well as with the nature of the interaction between the vicinal smectic layers and the boundary itself; hence,  leading to a fluctuation-induced pseudo-Casimir force between the bounding substrates, which we investigated in detail in this work. 

We use the standard statistical mechanical continuum field theory paradigm to calculate the fluctuation force,  resulting from thermally excited modes within a $\kB T$ energy range of their ground state. While, in general, the context of our calculation is quite close to the pseudo-Casimir interactions, we do explicitly take into account the smectic-A boundary interaction energy, assuming it has a general Rapini-Papoular form~\cite{RP}, and thus avoid the question of the proper boundary condition of the fluctuating displacement fields. In addition, our calculation is interesting also from the purely formal perspective, since it develops further the Casimir methodology of field Hamiltonians, involving higher order field derivatives, which were first used in the context of Casimir interactions for confined nematic polymers~\cite{dobnikar} and confined Brazovskii-type soft media~\cite{uchida}. The main difference between the Brazovskii-type soft media and the smectic-A liquid crystalline film is the absence of the bound states in the Fourier decomposition in the direction perpendicular to the bounding surfaces. This difference can be tracked to the sign of the first order derivative term in the field Hamiltonian Eq. \eqref{ham_b}.

The main features of our results may be outlined as follows. First, the fluctuations with vanishingly small values of both the azimuthal and normal derivatives of layer displacement field on the bounding substrates are found to give  a long-ranged intersubstrate attraction, which, at sufficiently large separations, is well-described by the universal expression given in Eq. \eqref{eq:pressure_0}. The calculated prefactor $C_0\simeq 1.76985$ differs from what is proposed in Ref. \cite{LC2}, where the prefactor is suggested to be of the form $\zeta (4)\simeq 1.08232$. The latter conjecture is indeed not supported by our numerical findings. Second, the interaction force (or pressure) exhibits a non-monotonic dependence on  the ratio $\lambda/\ell$ and shows a duality with respect to the magnitude of the surface anchoring energy. Third, the attractive pressure mediated between the bounding substrates is strengthened as  $a_0$,  the periodicity along the normal of the layers, is decreased, implying that denser smectic systems should show an increased pseudo-Casimir force. 

The non-monotonic behavior of the pseudo-Casimir interaction pressure on the surface anchoring coupling strength $\lambda/\ell$ leads to small interaction magnitude for intermediate values, that increases either for very large or very small values of the surface coupling strength.
The interaction forces obtained in the two limits of weak ($\lambda/\ell\rightarrow 0$) and strong anchoring ($\lambda/\ell\rightarrow \infty$) coincide, indicating a special case of a duality in this system, which resembles the behavior previously seen in the case of nematic liquid crystals 	 \cite{LC2,rudi1,karimi_thesis}.
	 This property can be understood by examining the boundary condition equations as follows. Formally, for a bulk free-energy density of the form
$f_b={1\over 2}a \phi^2+{1\over 2}b \dot \phi^2+{1\over2 }c \ddot \phi^2$ and  the surface energy density of the form $f_s=W({1\over 2}\eta\phi^2+{1\over 2}\zeta\dot\phi^2)$, as we deal with in this paper, the boundary equations at $y=d$ read $b\dot \phi-c\dddot\phi+W\eta \phi=0$ and $c\ddot\phi+W\zeta\dot\phi=0$. We note that the limit $W \rightarrow \infty $ corresponds to $\phi=\dot\phi=0$ on the substate. Then the partition function directly reads
${\cal Z}={\cal A}\propto 1/\sqrt{M}$ [see Eqs.~(\ref{partition_fuction}), (\ref{eq:A_h_s_eff}), and (\ref{F_beta})], where $M$ turns out to be equal to $2G\big|_{W\rightarrow 0,\infty}$ [Eq.~(\ref{G_limit})].  On the other hand, for $W=0$, the boundary condition equations lead to undetermined values for $\phi$ and $\dot\phi$. This is why to evaluate  ${\cal Z}$ we integrate (average) over all possible values of both fields. This averaging reasonably leads to the same result as for the strong anchoring condition.
\begin{acknowledgments}
F.K.P.H would like to thank Ali Qariarab who took part, under his M.Sc. thesis, in initial studies of the subject.  A.N. acknowledges partial support from the Associateship Scheme of The Abdus Salam International Centre for Theoretical Physics (Trieste, Italy), and thanks the University of Chinese Academy of Sciences and the Institute of Physics, Chinese Academy of Sciences, for their support and hospitality during a scientific visit leading to the completion of the final form of this paper. R.P. also acknowledges the support of the 1000-Talents Program of the Chinese Foreign Experts Bureau and the hospitality of the University of the Chinese Academy of Sciences.
\end{acknowledgments}


\end{document}